\documentclass[aps,prl,reprint,superscriptaddress]{revtex4-1}

\usepackage{xcolor}
\usepackage{booktabs, multirow} 
\usepackage{makecell}

\usepackage{graphicx}
\usepackage{hyperref}

\begin{document}


\title{Exciton nature of plasma phase transition \\in warm dense fluid hydrogen}


\author{I.~D.~Fedorov}
    \email[]{fedorov.id@mipt.ru}
    \affiliation{Joint Institute for High Temperatures of the Russian Academy of Sciences, Izhorskaya 13 Building 2, Moscow 125412, Russian Federation}
    \affiliation{Moscow Institute of Physics and Technologies (National Research University), Institutskij pereulok 9, Dolgoprudny Moscow region 141700, Russian Federation}
\author{V.~V.~Stegailov}
    \email[]{stegailov.vv@mipt.ru}
    \affiliation{Joint Institute for High Temperatures of the Russian Academy of Sciences, Izhorskaya 13 Building 2, Moscow 125412, Russian Federation}
    \affiliation{Moscow Institute of Physics and Technologies (National Research University), Institutskij pereulok 9, Dolgoprudny Moscow region 141700, Russian Federation}
    \affiliation{HSE University, Myasnitskaya ulitsa 20, Moscow 101000 Russian Federation}


\date{\today}

\begin{abstract}
The transition of a warm dense fluid hydrogen from insulator to a conducting state at pressures of the order of 20–300~GPa and temperatures of 1000–5000~K has been the subject of active scientific research over the past few decades. The use of various experimental methods has not yet led to obtaining reliable consistent results, and despite numerous attempts, theoretical methods have not yet been able to explain the existing discrepancies, as well as the microscopic nature of the mechanism for the transition of hydrogen fluid to a conducting state. In this work, we have discovered a key mechanism of the transition associated with formation and dissociation of exciton pairs, which allows to explain several stages of the transition of hydrogen from molecular state to plasma. This mechanism is able to give quantitative description of several experimental results as well as to resolve the discrepancies between different experimental studies.
\end{abstract}


\maketitle

\textit{Introduction.} The behavior of hydrogen under conditions of extreme compression (100-400~GPa) and high temperatures (1000-4000~K) has been studied for quite a long time~\cite{mcmahon2012properties,Utyuzh_2017, helled2020understanding, silveradiasreview2021, normansaitovreview2021} but its theoretical description still poses many unresolved questions~\cite{Liu13374, PhysRevE.104.045204, ishkhanyan2021van}. One of these problems is the transition of compressed hydrogen in both solid and fluid states into a conducting state (metallization of solid hydrogen was predicted a long time ago~\cite{wigner1935possibility}). Over the past few decades, a large number of experimental works have been carried out, both using shock compression~\cite{fortov2007phase, Knudson1455, mochalov2017quasi, Celliers2018} and using pulsed heating in diamond anvil cells~\cite{Loubeyre-etal-HighPressRes-2004, Dzyabura2013, ohta2015phase, goncharov2016, zaghoo2018striking, houtput2019finite_PRB, jiang2020spectroscopic}. At the same time, there is still no unambiguous theory that explains the experimental results as well as multiple differences in the observations of different groups.

Initially, the transition of fluid hydrogen from the molecular state to the atomic state was considered from the point of view of chemical models~\cite{Norman1968,Biberman1969,Lebowitz1969,Norman1970,Norman1970a,Ebeling1971,Kraeft1986,Saumon1991,Reinholz1995,Ebeling2003,Norman2006,Khomkin2013,Starostin2016,Ebeling2017}. The concept of plasma phase transition was introduced~\cite{Norman1968, FilinovNorman-PRA-1975, NormanSaitov-CPP-PPT50-2019}. Later, the insulator-to-metal transition concept has become widespread~\cite{Redmer2010book}. The most common \textit{ab initio} approaches are quantum molecular dynamics (QMD) based on density functional theory and quantum Monte Carlo methods (see ~\cite{Scandolo-PNAS2003, TamblynBonev-PhysRevLett.104.065702-2010, Morales-et-al-PhysRevLett.110.065702-2013, Mazolla2017, Knudson-Desjarlais-PhysRevLett.118.035501-2017, knudson2018_ftdft, KarasievDuftyTrickey-PhysRevLett.120.076401-2018, Ackland-etal-PhysRevB.100.134109-2019,  Norman2019, PhysRevB.101.195129} and ~\cite{DelaneyPierleoniCeperley-PhysRevLett.97.235702-2006,  Tubman-etal-PhysRevLett.115.045301-2015,Morales-etal-PNAS-2010, Pierleoni4953, MazzolaSorella-PhysRevLett.114.105701-2015, Mazzola-etal-PhysRevLett.120.025701-2018, Rillo9770,  PhysRevB.102.195133, tian2020first, PhysRevB.102.144108, PhysRevLett.126.225701}. At the same time, the plephora of methods cannot fully explain the key properties of the transition, such as a large isotope effect~\cite{zaghoo2018striking}, a large latent heat of transition~\cite{houtput2019finite_PRB}, as well as differences in the moments of the detected increase in absorption and reflectivity~\cite{Rillo9770}. 

In this work, we focus on two previously largely neglected phenomena: on the energy transfer from ions to electrons that is the driving force of thermal excitations leading to plasma phase transition and on the initial localized elementary processes of conducting phase formation. For this purpose, we deploy the methods within the density functional theory (DFT) that give the possibility to capture the corresponding non-abiabatic and non-equilibrium processes (in contrast to the finite temperature (FT) DFT method that is nowadays used routinely for warm dense matter calculations, e.g. see~\cite{Morales-et-al-PhysRevLett.110.065702-2013, Knudson-Desjarlais-PhysRevLett.118.035501-2017, knudson2018_ftdft, KarasievDuftyTrickey-PhysRevLett.120.076401-2018, Ackland-etal-PhysRevB.100.134109-2019, Norman2019, PhysRevB.101.195129}). 


Our previous study of the dynamics of individual electronic excitations~\cite{fedorovstegailov2020_prb} showed the possibility of the appearance of an localized excitons during the transition. The lifetimes of such excitons turned out to be on the order of 10~fs, which is much larger than a typical time step of \textit{ab initio} molecular dynamics. This fact casts doubt on the applicability of FT DFT, where the instantaneous nature of electronic transitions and thermalisation is assumed. In our further study~\cite{fedorovstegailov2021_jetp}, we showed the dissociation of such bound electron-hole ($e$-$h$) pairs. Excitons in liquids are not unusual, e.g. these states were discovered in Xe~\cite{beaglehole1965reflection,asaf1971wannier}. In this work, we extend our preliminary studies to the mechanism of the plasma phase transition in warm dense hydrogen and compare the results with experimental data.  

\textit{Model and calculation details.} For modelling of the first singlet excited electronic state (S1) of the system within DFT, we use the Restricted Open-Shell Kohn-Sham (ROKS) method~\cite{roks1998,cpmd2002}. 
ROKS uses the orbital representation of the many-electron wave function and therefore allows the analysis of singly occupied molecular orbitals (SOMOs) by calculation of the corresponding Wannier centers (the SOMO-1 orbital with lower energy is interpreted as a hole ($h$) and the higher SOMO-2 orbital is interpreted as an electron ($e$)~\cite{schwermann2020exciton}). Together they describe an $e$-$h$ pair or an exciton.


\begin{figure}
\includegraphics[width=8.6cm]{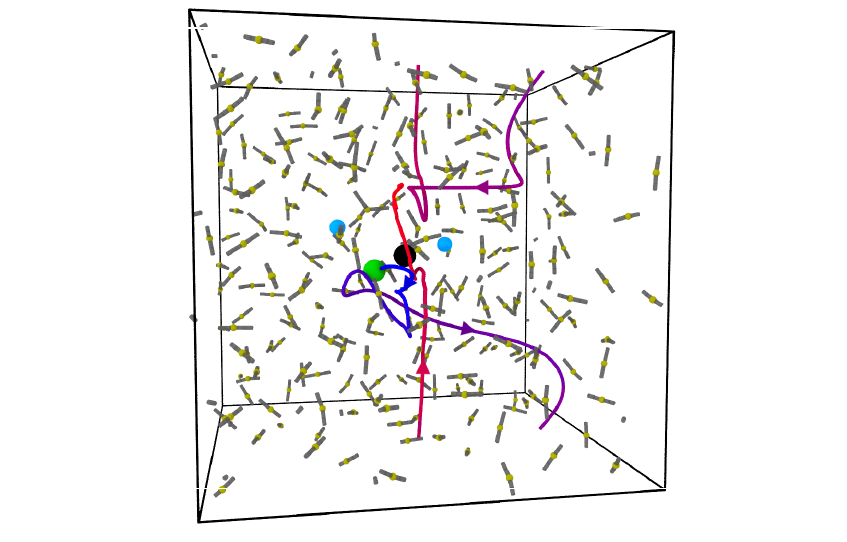}
\caption{The system of 480 hydrogen atoms (0.8~g/cc, 1400~K). A gradient color line shows the 30~fs trajectory of the Wannier center of SOMO-2 ($e$) when the simulation box is centered each time under periodic boundary conditions with respect to the Wannier center of SOMO-1 ($h$). 
 (see more in Supplementary Materials). \label{fig:rho_08_1400_visual}}
\end{figure}

The calculations are performed using CPMD~\cite{CPMD-v3.17} with the BLYP exchange-correlation functional. The QMD step is 4 a.u.
480 hydrogen atoms is considered in the density range 0.6-1.0~g/cc at temperatures 600-3000~K corresponding to pressures 60-240~GPa.
A plane wave basis with a cutoff of 70~Ry is used. The summation over the Brillouin zone is replaced by the value at the $\Gamma$-point. The initial configurations for each density and temperature are obtained by Car-Parrinello MD in the electron ground state (S0) using the Nose-Hoover thermostat.
Then, Born-Oppenheimer MD in S1 without thermostat is carried out within ROKS at times of at least 1~ps with the derivation of Wannier centers at each MD step (see Fig.~\ref{fig:rho_08_1400_visual}). 

Due to the technical difficulties of pressure calculations in CPMD, we use VASP~\cite{kresse1996efficient} for pressure calculations in the ground state (S0) (see Supplementary Materials).

\textit{QMD results.} Previously~\cite{fedorovstegailov2020_prb}, we have shown the mechanism of spontaneous non-adiabatic energy transfer from molecular kinetic energy to excitons. The probability of the formation of such excitons increases with temperature. In this work, we analyze the properties of a model of a single exciton at different temperatures and densities and determine the conditions of exciton dissociation when energy transfer from ions to electrons becomes irreversible leading to the plasma phase transition.

\begin{figure}
\includegraphics[width=8.6cm]{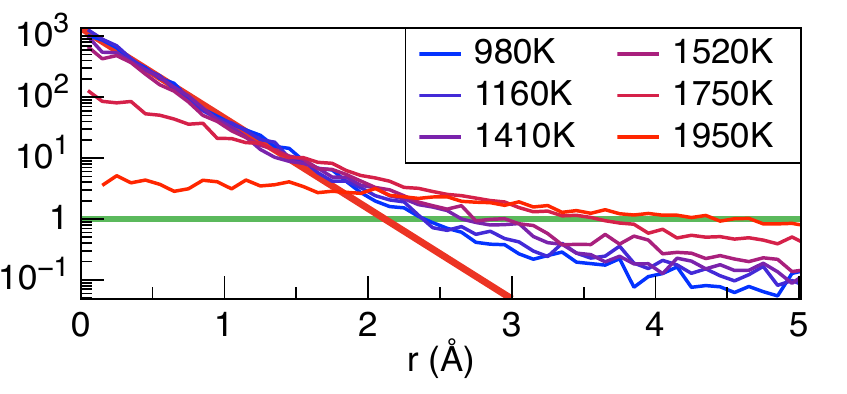}
\caption{RDFs for the Wannier centers of SOMO-1 ($h$) and SOMO-2 ($e$) orbitals for 0.8~g/cc (other densities are presented in Supplementary Materials). 
The red thick line reflects the $\exp(-r/a)$ distribution that holds at short distances and lower temperatures. 
\label{fig:rdf_somo1_somo2}}
\end{figure}

As can be seen from the visualization of the system in Fig.~\ref{fig:rho_08_1400_visual}, there are protons not bound to any Wannier center near the considered $e$-$h$ pair (the SOMO-1 and SOMO-2 centers, respectively). This is a regular picture (see the animation in Supplementary Materials). It is worth noting that such ``free'' protons collide quite often with hydrogen molecules and replace their protons. Moreover, as can be seen from the displayed trajectory of the electron relative to the hole (Fig.~\ref{fig:rho_08_1400_visual}), most of the time the $e$-$h$ distance is quite small. The analysis of such distances via the $e$-$h$ radial distribution function (RDF) (Fig.~\ref{fig:rdf_somo1_somo2}) shows that, up to certain temperatures, the initial parts of RDFs correspond to the $g(r) \sim \exp(-r/a)$ dependence that is a well-known behavior of charges in a Coulomb field. The tails of RDFs deviate from exponential dependence due to the fluid state of the system and high temperatures, which undoubtedly affect the electron and hole mobility.

\begin{figure}
\includegraphics[width=8.6cm]{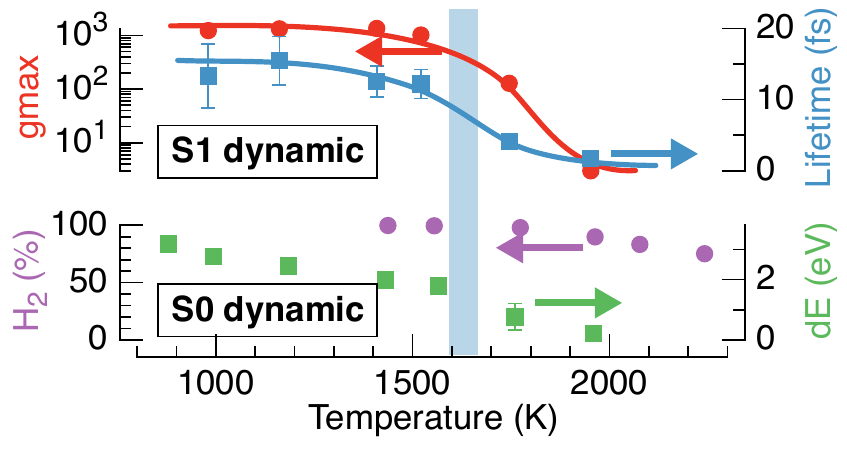}
\caption{The upper part (the data from S1 MD): the temperature dependence of the RDF maxima (red line with circles) and the lifetimes of bound excitons (blue line with squares). 
The lower part (the data from S0 MD): the concentration of molecular hydrogen (purple circles) and the S0$\to$S1 vertical excitation energy (green squares). 
All data are given for a density of 0.8~g/cc (the data for  0.6 and 1.0~g/cc are in Supplementary Materials). The temperature range of exciton dissociation is shown by a blue bar ($\tau_{dis}\sim 7-9$~fs).\label{fig:all_in_one}}
\end{figure}

The red line in Fig.~\ref{fig:all_in_one} shows the dependence of the height of the RDF maximum $g_{max}=g(r=0)$ on temperature (determined by the $\exp(-r/a)$ approximation of RDFs at short distances). One can notice a sharp drop with increasing temperature that indicates the beginning of the dissociation of excitons. Such a dissociation leads to the fact that the reverse transitions to the ground electronic state S0 become unlikely, due to the spatial separation of the $e$-$h$ pair. 

Another important characteristic of an exciton is the distance at which RDF crosses the value of 1 (the green line in Fig.~\ref{fig:rho_08_1400_visual}). This distance corresponds to the typical size of a $e$-$h$ pair at a given temperature (the size effect was checked with a system of 240 H atoms, see Supplementary Materials). Using this distance as a threshold, we obtain the average waiting time of dissociation during which a $e$-$h$ pair is in a bound state $\tau_{dis}$ (see the blue line in Fig.~\ref{fig:all_in_one}). The obtained dependence of lifetime $\tau_{dis}$ correlates quite well with the behavior of $g_{max}$ and demonstrates a sharp decrease with increasing temperature as well. 

The lower threshold for the values of exciton lifetimes that determines the irreversible dissociation of the $e$-$h$ pair is the average vibrational period of H$_2$ molecules $\tau_{vib} \sim 7-9$~fs. The temperatures when exciton lifetime becomes similar or lower than the vibrational period of hydrogen molecules are the temperatures when the energy transfer from ions to electronic excitations becomes irreversible. We consider this process as the first stage of the plasma phase transition (see below). 


The analysis of the averaged molecular composition at a given temperature is carried out by sorting all the positions of protons along the equilibrium S0 QMD trajectory into H$_2$ pairs with distances less than 1.0~{\AA} living longer than 10~fs. The data in Fig.~\ref{fig:all_in_one} (the purple circles) show that at the exciton dissociation temperatures (marked by the blue bar) the system has molecular composition.

The $dE=E(S1)-E(S0)$ values averaged over 250~fs along equilibrium QMD trajectories are displayed in Fig.~\ref{fig:all_in_one} as green squares. As can be seen from the graph, at exciton dissociation temperatures, the values of $dE$ is not lower than 1 eV. Therefore, we can conclude that the fluid H$_2$ is not metallic at the beginning of transition.


\begin{figure*}[ht!]
\includegraphics[width=17.2cm]{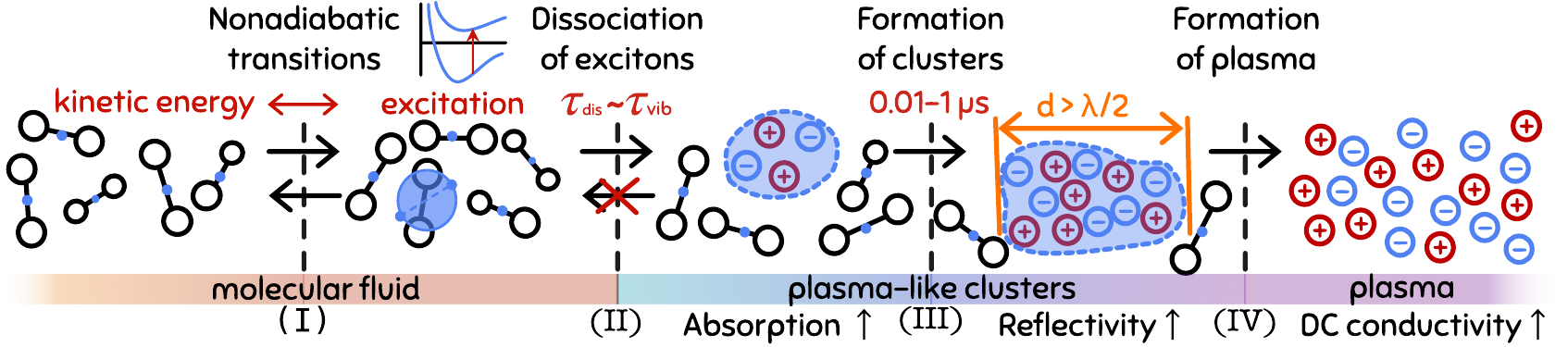}
\caption{The scheme of the transition mechanism proposed. Non-adiabatic transitions (I) have been considered previously~\cite{fedorovstegailov2020_prb}. Dissociation of Frenkel excitons is modelled in this work. Later stages are hypothesised and correlated with experimental data.
  \label{fig:sketch}}
\end{figure*}

\textit{Transition mechanism.} Thus, we propose the following mechanism for the transition of a dense fluid molecular hydrogen to plasma at heating (Fig.~\ref{fig:sketch}).

I. (formation of short-lived Frenkel excitons via non-adiabatic energy transfer). At the beginning, we have fluid molecular hydrogen with molecules moving on the potential energy surface of the ground state (S0). Then, as the temperature rises, the number of non-adiabatic transitions of electron subsystem from S0 to S1 increases. Each transition is a result of the atomic kinetic energy transfer to an electronic excitation and leads to the formation of an exciton. We assume that in experiments independent excitons can form simultaneously in different parts of the macroscopic fluid H$_2$ sample.  The probability of such non-adiabatic excitations depends on atomic velocities that contributes to a \textit{large H/D isotope effect} for this transition~\cite{zaghoo2018striking} (as discussed in~\cite{fedorovstegailov2020_prb}). The average $e$-$h$ distance in these excitons is about the size of an H$_2$ molecule that is why we can call them Frenkel excitons. Most $e$-$h$ pairs formed recombine spontaneously via internal conversion S1$\to$S0 after $\tau_{vib}$~\cite{fedorovstegailov2020_prb}.  

\begin{figure}[ht]
\includegraphics[width=8.6cm]{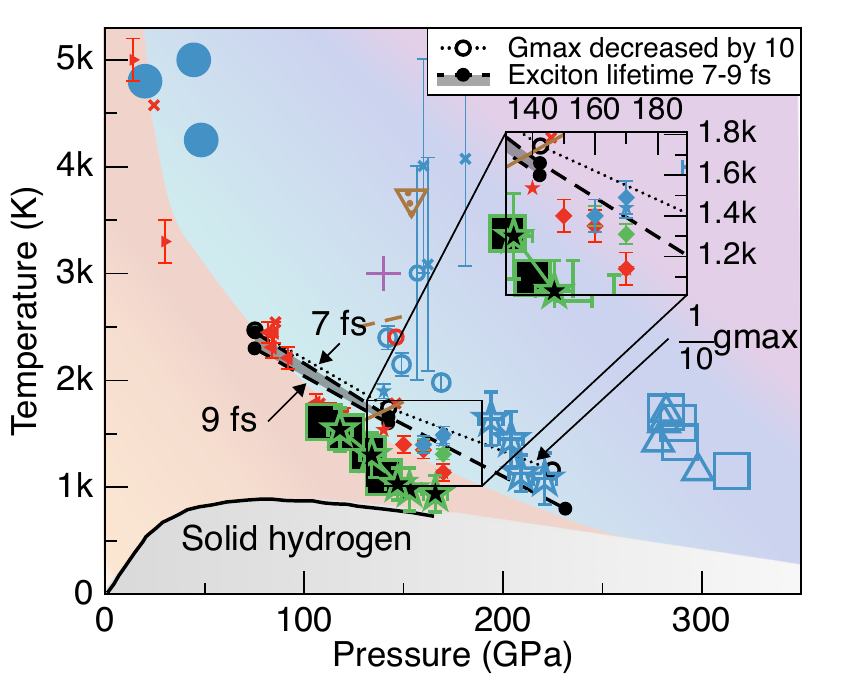}
\caption{T-P phase diagram. The grey area between two dashed lines with black circles indicates that the Frenkel exciton lifetime $\tau_{dis}$ becomes of the order of the oscillation period of the hydrogen molecule ($\tau_{vib} = 7–9$~fs). The ten-fold decrease of $e$-$h$ RDF $g_{max}$ is shown by the dotted line. The experimental marks attributed to the plasma phase transition are shown: the temperature plateau (red points), the increase in optical absorption (green points), the increase in optical reflectivity (blue points), the increase in DC conductivity (purple points). The filled symbols correspond to H, the open symbols correspond to D. Larger symbols correspond to experiments with higher heating/compression rates (detailed descriptions of the points with brief descriptions of the experimental works are presented in Supplementary Materials).\label{fig:pt}}
\end{figure}

II. (formation of Wannier-Mott excitons). At higher temperatures, dissociation of Frenkel excitons becomes very fast with their average lifetime smaller than the average vibrational period of H$_2$ molecules $\tau_{dis} \lesssim \tau_{vib}$. Each dissociated Frenkel exciton becomes less delocalized (we see no pronounces maximum on $g(r)$) but still bound to protons via electrostatic interactions. This stage is currently beyond our capabilities for direct QMD calculations. That is why we describe this stage and the next stages only qualitatevly. Such dissociated Frenkel excitons can be called Wannier-Mott excitons. This exciton dissociation leads to the irreversible energy transfer from kinetic energy of molecules to electronic excitations. Each dissociation event consumes about $dE \sim 1-2$~eV, which are taken from the kinetic energy of the ions. 
It explains the ``latent heat of transition'' of the order of 1~eV/atom that comes from the analysis of experimental data~\cite{houtput2019finite_PRB} (in contrast to the hydrogen dissociation model that predicts this value to be about 0.05 eV/atom only~\cite{Morales-etal-PNAS-2010}. This exciton dissociation leads to a \textit{plateau in the dependence of the peak temperature} of fluid H$_2$ on the absorbed laser energy observed in many works~\cite{Dzyabura2013, ohta2015phase, goncharov2016, zaghoo2016evidence, Zaghoo11873,jiang2020spectroscopic} (see the red points in Fig.~\ref{fig:pt}). Formation of many Wannier-Mott excitons in a macroscopic fluid H$_2$ sample explains the \textit{rise of absorption of radiation} by fluid H$_2$~\cite{ zaghoo2016evidence, Knudson1455, Celliers2018} (see the green points in Fig.~\ref{fig:pt}). The threshold exciton dissociation temperatures (when $\tau_{dis} \approx \tau_{vib}$) depend not on the kinetic energy of molecules but on their velocities (in contrast to molecular dissociation that depends on energy of molecular vibrations). That is why the threshold temperatures for D are expected to be about 2 times higher than for H. We can consider this mechanism as another contribution to the detected \textit{large H/D isotope effect} for this transition~\cite{zaghoo2018striking}.

III. (formation of reflective clusters from Wannier-Mott excitons). The next stage of the plasma phase transition is the formation of localized plasma-like clusters from single Wannier-Mott excitons. As soon as the average size of such clusters becomes comparable with the wavelength of probing radiation, \textit{the rise of optical reflectivity} by fluid H$_2$ is detected~\cite{Loubeyre-etal-HighPressRes-2004, loubeyre2012,zaghoo2016evidence, zaghoo2018striking, jiang2020spectroscopic, Knudson1455,Celliers2018} (see the blue points in Fig.~\ref{fig:pt}). The electrons in such plasma clusters is not absolutely free but bound to the corresponding clusters (similarly to the collective excitations discussed in~\cite{LankinNorman-CPP-2009}). In the single-particle electronic energy spectrum for Kohn-Sham orbitals, their energies are lower than the continuum level but could be higher than the SOMO-2 energy level.


IV. (formation of plasma). Eventually, the plasma is formed that leads to the observation of DC conductivity~\cite{weir1996metallization, fortov2003pressure, fortov2007phase} (see the purple points in Fig.~\ref{fig:pt}).

\textit{Discussion of experiments.} Fig.~\ref{fig:pt} shows the T-P diagram of hydrogen with the regions of plasma, fluid and solid hydrogen. Various transient processes observed experimentally are assigned to the plasma phase transition considered. The (P,T) points on Fig.~\ref{fig:pt} show the threshold conditions interpreted as the plasma phase transition points.  Experimental studies can be divided into shock wave (SW) and diamond anvil cell (DAC) experiments.

According to the transition mechanism proposed in this work, during heating of fluid molecular hydrogen we can distinguish the following phenomena detected by various techniques: (1) a temperature plateau (shown as red dots) that is explained by non-adiabatic excitations of a Frenkel excitons and their irreversible dissociation that transfers $\sim$1-2~eV per event from kinetic energy of molecules to electronic excitations, (2) the rise of optical absorption (green dots) that is explained by formation of separated Wannier-Mott excitons that can absorb light but do not reflect, (3) the rise of optical reflection (blue dots) that is explained by the formation of excition (plasma-like) clusters with the size $d > \lambda/2$, where $\lambda$ is the probing light wavelength, and (4) an increase in DC conductivity (purple dots) due to plasma formation. 

Comparison of our QMD results with experimental data shows that our (P,T) points obtained with BLYP xc-functional and ROKS as the thresholds of Frenkel excitons dissociation agree quite well with the DAC results of Ohta et al.~\cite{ohta2015phase} and Zaghoo et al.~\cite{zaghoo2016evidence,Zaghoo11873}. On the one side, BLYP and ROKS should be considered as quite generic approximations to the electronic structure of the dense molecular hydrogen in the first excited state (the dependence of the results on the choice of xc-functional was shown previously~\cite{Knudson-Desjarlais-PhysRevLett.118.035501-2017}). On another side, such simple approximations as BLYP and ROKS could be indeed quite accurate for such a simple system as dense molecular hydrogen (no strong correlations and no pronounced dispersion interactions). Therefore, good agreement with experiments could reflect the fact that in DAC experiments the threshold (P,T) states correspond to the Frenkel exciton dissociation conditions. Relatively slow heating in DAC experiments~\cite{zaghoo2016evidence,Zaghoo11873} within the framework of the transition mechanism proposed can explain why the temperature plateau, the onset of absorption and the rise of reflectivity are not separated in time. Indeed, reflective plasma clusters can form on the microsecond timescale of these DAC experiments.

The corresponding SW experiments are about one order of magnitude faster than the DAC experiments (see Supplementary Materials). The difference in the moments when absorption and reflectivity in SW studies are detected can be explained by the kinetics of plasma-like clusters formation. Before the reflective clusters of the size $d > \lambda/2$ are formed, the system has been already driven to higher temperatures and pressures. Different kinetics of plasma-like clusters formation can explain different (P,T) threshold values when the rise of reflectivity is observed in different experiments (e.g. Z-machine~\cite{Knudson1455} vs NIF~\cite{Celliers2018}). In SW experiments temperatures are not measured directly but calculated using an equation of state (EOS) that bring another source of possible inaccuracies. In some SW experiments~\cite{fortov2007phase, mochalov2017quasi}, a density jump is observed during compression (Fig.~\ref{fig:pt}, brown dots). It is noteworthy that one of EOS variants places this jump near the DAC results and near our QMD results for exciton dissociation. Moreover, some of the high-temperature points can explained by the fact that they were observed in DAC on cooling~\cite{jiang2020spectroscopic}, when the kinetics of plasma recombination is expected to differ strongly from the kinetics of plasma formation on heating (for example, in SW studies~\cite{Knudson1455} a hysteresis of optical reflection was observed).


\textit{Conclusions.} The new mechanism of plasma phase transition in warm dense fluid hydrogen has been proposed based on the analysis of formation and growth of localized electronic excitations during heating. Kinetics of these processes explains both the existing discrepancies in experimental data and the key experimental observations: the temperature plateaus and the large detected ``latent heat of transition'', the large isotope effect, the variations in the moments of absorption and reflectivity rise. The key threshold process of the proposed mechanism is the dissociation of excitons. QMD calculations with the BLYP exchange-correlation functional and the ROKS method of S1 electron dynamics give the threshold temperatures and pressures of the transition that agree quite well with the appropriate experimental data.

\begin{acknowledgments}
We thank Nikos Doltsinis and Genri Norman for useful discussions. This research was supported by The Ministry of Science and Higher Education of the Russian Federation (Agreement with Joint Institute for High Temperatures RAS No.~075-15-2020-785 dated September 23, 2020). The authors acknowledge the Supercomputer Centre of JIHT RAS and the Supercomputer Centre of MIPT. This research was supported in part through computational resources of HPC facilities at NRU HSE.
\end{acknowledgments}

%

\end{document}